# Periodic orbits in the gravity field of a fixed homogeneous cube

Xiaodong Liu[1], Hexi Baoyin[2], and Xingrui Ma[3]

*School of Aerospace, Tsinghua University, Beijing 100084, China*

Email:   *liu-xd08@mails.tsinghua.edu.cn*; *baoyin@tsinghua.edu.cn*; *maxr@spacechina.com*

**Abstract**

In the current study, the existence of periodic orbits around a fixed homogeneous cube is investigated, and the results have powerful implications for examining periodic orbits around non-spherical celestial bodies. In the two different types of symmetry planes of the fixed cube, periodic orbits are obtained using the method of the Poincaré surface of section. While in general positions, periodic orbits are found by the homotopy method. The results show that periodic orbits exist extensively in symmetry planes of the fixed cube, and also exist near asymmetry planes that contain the regular Hex cross section. The stability of these periodic orbits is determined on the basis of the eigenvalues of the monodromy matrix. This paper proves that the homotopy method is effective to find periodic orbits in the gravity field of the cube, which provides a new thought of searching for periodic orbits around non-spherical celestial bodies. The investigation of orbits around the cube could be considered as the first step of the complicated cases, and helps to understand the dynamics of orbits around bodies with complicated shapes. The work is an extension of the previous research work about the dynamics of orbits around some simple shaped bodies, including a straight segment, a circular ring, an annulus disk, and simple planar plates.

---

1  PhD candidate, School of Aerospace, Tsinghua University
2  Associate Professor, School of Aerospace, Tsinghua University
3  Professor, School of Aerospace, Tsinghua University

## 1. Introduction

Previous literature has mentioned that to study orbits around the Platonic solids might be fruitful (Werner 1994). The investigation of orbits around these simple shaped bodies could be considered as the first step of the complicated cases, and helps to understand the dynamics of orbits around bodies with complicated shapes. The cube, which is a simple one of the five Platonic solids, is the subject of this paper. The dynamics of a particle orbiting around a fixed homogeneous cube are complex despite the cube's simple shape, and several types of periodic orbits are found. Some relevant research was made in the past. It was once proved that there exist ring-type bounded planar motions in an isolated system consisting of a homogeneous cube and a massive point particle (Michalodimitrakis and Bozis 1985). In our previous research, equilibria of motion around a rotating cube were derived, and invariant manifolds connecting periodic orbits around the equilibria were calculated (Liu et al. 2011).

A new thought of searching for periodic orbits around non-spherical bodies is provided in this paper. The cube, as a simple case of non-spherical bodies, is taken as an example to prove the effectiveness of this new thought, i.e. the homotopy method. In artificial satellite theory, periodic orbits around non-spherical celestial bodies are of special interest. Periodic orbit families about asteroid 4769 Castalia were computed using the Poincaré maps and Newton–Raphson iteration (Scheeres et al. 1996). For asteroid 4179 Toutatis in a non-principal-axis rotation state, periodic orbits were found by using the averaged equations of motion to generate the initial conditions and computing the state transition matrix to adjust them (Scheeres et al. 1998). The computation of periodic orbits around asteroid 433 Eros was made with the application of the Poincaré maps and the monodromy matrix (Scheeres et al. 2000). In this paper, the homotopy method, which does not make use of symmetrical characteristic, is proposed to find periodic orbits. It is proved effective for the case of the cube in Sect. 5, and is also applicable to search for periodic orbits around other non-spherical bodies. Although the homotopy method has been introduced for many years, so far there is little literature about the application of the homotopy method to search for periodic orbits around celestial bodies.

This study can be considered as an extension of the previous researches about the cases of some simple shaped massive bodies with backgrounds in celestial mechanics. For fixed massive bodies, periodic orbits and other dynamics characteristics around a straight segment (Riaguas et al. 1999; Arribas and Elipe 2001), a solid ring (Broucke and Elipe 2005; Azevêdo et al. 2007; Azevêdo and Ontaneda 2007; Fukushima 2010), a homogeneous annulus disk (Alberti and Vidal 2007; Fukushima 2010), and simple planar plates including square and triangular plates (Blesa 2006) were investigated. The dynamics around a rotating segment were extensively studied (Riaguas et al. 2001; Elipe and Riaguas 2003; Gutiérrez–Romero et al. 2004; Palacián et al. 2006). In these previous studies, the simple shaped bodies are limited to one- or two-dimensional cases. In contrast, the current study extends the simple shaped bodies to three dimensions.

Moreover, this study contributes to the investigation of irregular shaped celestial bodies that can be divided into a set of the cubes. In this way, the potential of the celestial bodies can be obtained by summing the potentials of all cubes. To get a clear understanding of the dynamics of orbits around the cube, which can be taken as the basic unit, provides valuable information and help to investigate the irregular shaped celestial bodies. The traditional method to represent the gravity field of celestial bodies is to expand the potential into series of spherical harmonics, which is effective for spheroid-like bodies. However, for irregular shaped celestial bodies, such as asteroids and comet nuclei, this method may fail to converge when within the circumscribing sphere centered at the harmonic expansion center and surrounding the shape. To remedy drawbacks of the harmonic expansion, the mascon approximation (Geissler et al. 1996; Scheeres et al. 1998) and the polyhedral approach (Werner 1994; Werner and Scheeres 1997) were proposed. Based on the polyhedral approach, irregular shaped celestial bodies can also be divided into a finite number of simple-shaped basic units, such as cubes or tetrahedral (Werner and Scheeres 1997).

The gravity field of 433 Eros was once modeled as the summation of gravity fields of a set of tetrahedra (Scheeres et al. 2000).

In this study, the dynamics of orbits around a fixed homogeneous cube are considered. The potential of the fixed cube is calculated by the polyhedral approach. Periodic orbits in the symmetry planes are derived by computing the Poincaré surface of section. While in General Positions of the cube, periodic orbits are found via the homotopy method. It is proved that periodic orbits exist extensively in symmetry planes of the fixed cube, and also exist near asymmetry planes that contain the regular Hex cross section. The stability of these periodic orbits is computed.

## 2. The gravitational potential of the homogeneous cube

Given a particle $P$ with infinitesimal mass located outside of the fixed homogeneous cube with edge length $2a$ and constant mass density $\sigma$, the potential at a certain point $P$ can be written as an integral over the volume of the body

$$U = G\sigma \iiint_V \frac{1}{r} \mathrm{d}V, \qquad (1)$$

where $G$ is the gravitational constant, and $r$ is the distance from the point $P$ to the differential mass of the cube. Establish a right-handed Cartesian coordinate system $Oxyz$ with the origin $O$ located at the center of the fixed cube, and the three coordinate axes coinciding with the symmetry axes of the cube.

The polyhedral approach (Werner and Scheeres 1997) is used to calculate the potential of the cube. The result is provided in the Appendix. Compared to the finite straight segment's potential (Riaguas et al. 1999), the solid circular ring' potential (Broucke and Elipe 2005), and the homogeneous annulus disk' potential (Azevêdo and Ontaneda 2007), the cubic potential is much more complex. It is evident that the potential is symmetrical with respect to the planes passing through the center, and parallel to the face as well as to the diagonal planes of the cube.

In this study, scaling is performed such that the half-length of the edge $a$ is the unit of length, and it is assumed that $G\sigma = 1$ in order to simplify the computation. The motion equation of the particle in the gravity field of the fixed cube can be obtained

$$\ddot{\mathbf{r}} = \nabla U .\qquad(2)$$

## 3. Periodic orbits in the symmetry planes parallel to the face of the fixed cube

It can be seen in Fig. 1 that $xy$-plane is one symmetry plane parallel to the face of the cube. Given the particle $P$ that is initially in an orbit tangent to the $xy$-plane, then the motion of $P$ will stay in the $xy$-plane due to the symmetrical characteristic of the potential. It is evident that the conservative dynamics system only has two degrees of freedom, and the motion of the particle in $z$ direction is zero all the time.

In this section, the method of the Poincaré surface of section is used to find periodic orbits. By integrating many orbits with the same value of the Energy constant, the Poincaré surface of section can be obtained by plotting $(x, \dot{x})$ whenever the particle crosses the $x$-axis.

The Poincaré surfaces of section for two different energy levels $E=-1.2$ and $E=-1.0$ are shown in Fig. 2. The case of $E=-1.2$ is used in this study. One central point that corresponds to the stable periodic orbit in the $xy$-plane can be noticed.

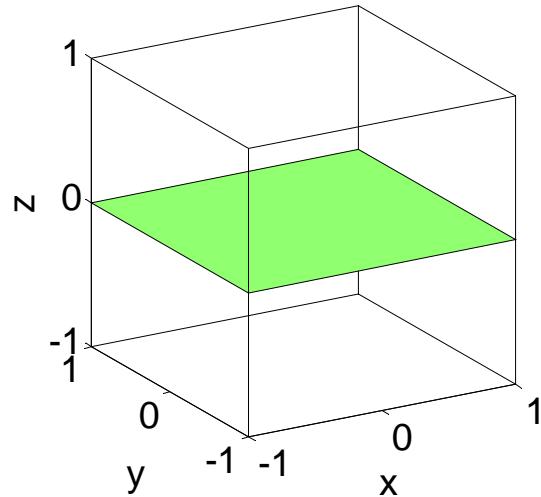

**Fig. 1** One symmetry plane parallel to the face of the cube: $xy$-plane.

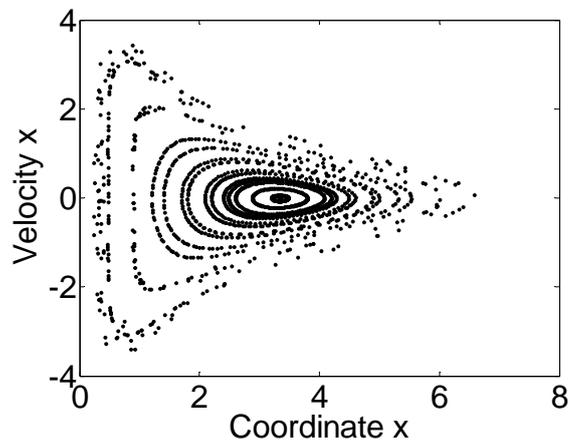

(**a**)

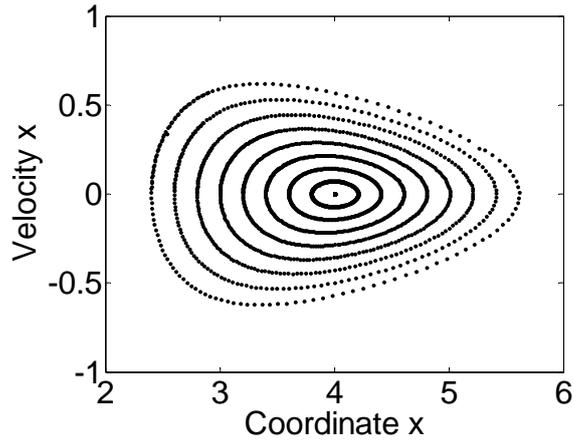

(**b**)

**Fig. 2** Poincaré surface of section for orbits in the *xy*-plane. (**a**) *E*=-1.2. (**b**) *E*=-1.0.

From the Poincaré surface of section shown in Fig. 2(a), the good initial condition for the periodic orbit can be obtained

$$x_0 = 3.34, \dot{y}_0 = 1.543321002314535. \tag{3}$$

The orbit with this initial condition of about 100 revolutions is shown in Fig. 3. It can be seen that the orbit is periodic, and symmetric with respect to both the *x*-axis and the *y*-axis.

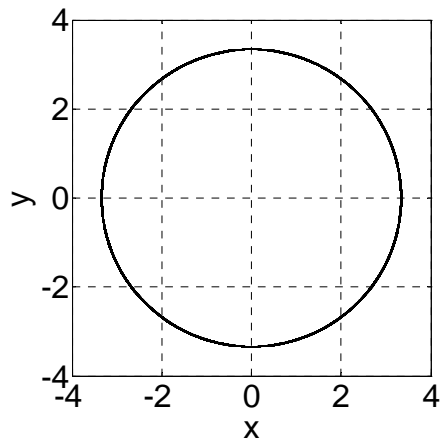

**Fig. 3** Periodic orbit in the *xy*-plane for *E*=-1.2.

## 4. Periodic orbits in the diagonal plane of the fixed cube

In the diagonal plane of the fixed cubic central body, the existence of the periodic orbits can also be proved. It is evident in Fig. 4 that the diagonal plane is also the symmetry plane of the fixed cube.

Another right-handed Cartesian coordinate system $Ox_1y_1z_1$ is established with three coordinate axes $\mathbf{i}_1$, $\mathbf{j}_1$, $\mathbf{k}_1$ parallel to $\mathbf{i}-\mathbf{k}$, $\mathbf{j}$, $\mathbf{i}+\mathbf{k}$, respectively. Given the particle $P$ that is initially in an orbit tangent to the diagonal plane, then the motion of $P$ will stay in this diagonal plane due to the symmetric characteristic of the potential. It is evident that the conservative dynamics system also only has two degrees of freedom, and the motion in $z_1$ direction is zero all the time.

Following the same process as Sect. 3, periodic orbits in the diagonal plane can also be found. The Poincaré surface of section for energy $E=-1.2$ is shown in Fig. 5. One central point that corresponds to the stable periodic orbit in the diagonal plane can be noticed.

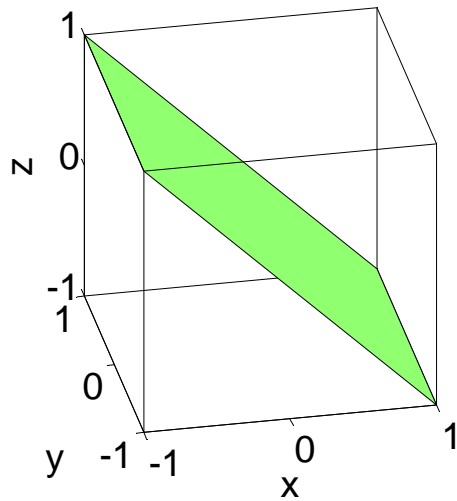

**Fig. 4** One diagonal plane of the cube.

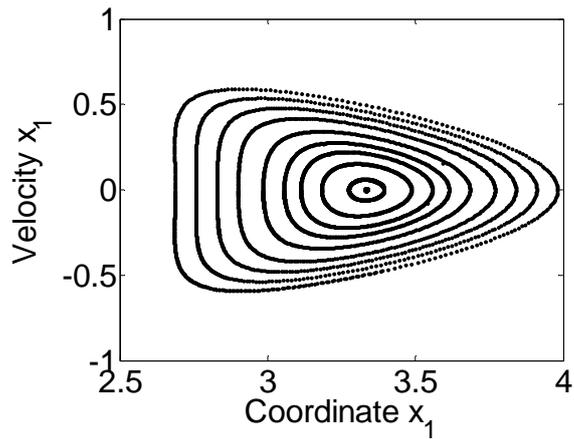

**Fig. 5** Poincaré surface of section for orbits in the $x_1y_1$-plane with $E$=-1.2.

From the Poincaré surface of section shown in Fig. 5, the good initial condition for the periodic orbit can be obtained

$$x_{10} = 3.337544007200505,\ \dot{y}_{10} = 1.547808464792009 \ . \tag{4}$$

The orbit with this initial condition of about 100 revolutions is shown in Fig. 6. It shows that this orbit is periodic, and is symmetric with respect to both the $x_1$-axis and $y_1$-axis.

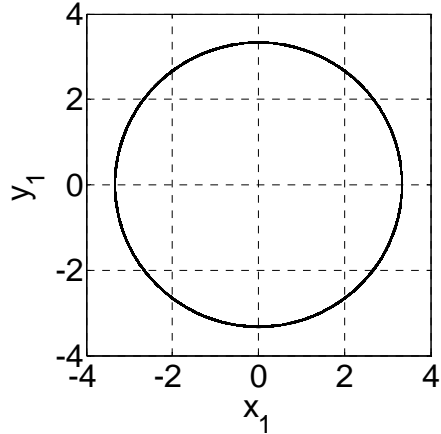

**Fig. 6** Periodic orbit in the $x_1y_1$-plane for $E=-1.2$.

**5. Searching of periodic orbits in general positions of the fixed cube**

In the asymmetry planes of the fixed cube, searching of periodic orbits is much more difficult than the symmetry cases because of the complexity of the potential of the fixed cube. The main difficulties come from the fact that the shape of the cube is not as smooth as that of the sphere.

In this section, the process involving with the homotopy method and Gauss-Newton algorithms is applied to overcome these difficulties. The homotopy method is a type of topology approach. A certain homotopy owns the right properties connecting the two problems, and the process of the homotopy method is to solve a more difficult one by starting from an easier one.

Let $f$ and $g$ be two mathematical objects from a space $X$ to a space $Y$. If there exists a continuous map from $X \times [0,1] \to Y$ such that $H(x,0) = f(x)$ and $H(x,1) = g(x)$, then the two mathematical are said to be homotopic.

In this study, the searching of periodic orbits of the fixed cube leads to a difficult problem, while searching for periodic orbits of the fixed sphere is much easier. It is obvious that the shape of the cube and the shape of the sphere are homotopic, so are the potential of the fixed cube and the potential of the fixed sphere. Therefore, a homotopy can be defined to connect the potentials of the fixed cube and the fixed sphere

$$H(\varepsilon) = \varepsilon U_{cube} + (1-\varepsilon) U_{sphere}, \tag{5}$$

where $U_{cube}$ is the potential of the fixed cube; $U_{sphere}$ is the potential of the fixed sphere. It can be easily derived that

$$H(0) = U_{sphere}, \quad H(1) = U_{cube}. \tag{6}$$

So the homotopy links the sphere problem (when $\varepsilon = 0$) and the cube problem (when $\varepsilon = 1$).

The homotopy process is to start at $\varepsilon = 0$, and increase $\varepsilon$ up to 1. The condition for $\varepsilon = 0$, corresponding to the sphere problem, is just Kepler problem, and much more regular. The initial guess for periodic orbits in this case is chosen as Kepler orbit elements. In this problem, $\varepsilon$ is increased from 0 up to 1 by the step size 0.01. For each step, the initial value of the iteration is taken as the periodic condition at the preceding step, and is integrated over one period $T$; the periodic condition for this step is obtained by Gauss-Newton algorithms and the process is iterated until the mismatch between the position and velocity at the initial time $t_0$ and the position and velocity at the final time $t_0+T$ is less than $10^{-8}$. Finally, when the value of $\varepsilon$ reaches 1, the condition for periodic orbits of the fixed cube can be obtained.

The stability of these periodic orbits can be determined by the eigenvalues of the monodromy matrix. The system is stable only if all eigenvalues of the monodromy matrix are located on the unit circle. The states of stability of periodic orbits appear in the eleventh column of Table 1, and each orbit is denoted by $U$ (unstable) or $S$ (stable).

Several initial conditions in the sphere problem are selected in order to obtain periodic orbits in the cube problem. The inclinations of these initial conditions are free parameters different from $0°$ to $90°$, and the initial semimajor axes $a$ are 4 or 5; while the other four Keplerian orbits elements of the these conditions are the same, with eccentricity $e=0$, right ascension of the ascending node $\Omega=20°$, argument of pericentre $\omega=0°$, and true anomaly $f=30°$. Via the homotopy method, each initial condition in the sphere problem leads to a periodic orbit around the cube. These periodic orbits around the cube can be divided into three different families: orbits in the planes parallel to the face of the cube, orbits in the diagonal plane, and orbits in the asymmetry planes that contain the regular Hex cross section (One regular Hex cross section of the cube is shown in Fig. 7). These families of about 100 revolutions are shown in Figs. 8, 9 and 10. All the necessary data of these periodic orbits are given in Table 1.

**Table 1** Data of periodic orbits exposed in Sect. 5.

| Orbit | Family | $a$ | $x_0$ | $y_0$ | $z_0$ | $v_{x0}$ | $v_{y0}$ | $v_{z0}$ | $T$ (s) | Stability |
|---|---|---|---|---|---|---|---|---|---|---|
| 1 | A | 4 | 3.2367087394 | 3.8120772276 | 0 | -0.9640378049 | 0.8184399380 | 0 | 24.8498127188 | $S$ |
| 2 | A | 5 | 3.3076964041 | 3.7588046209 | 0 | -0.9488195578 | 0.8348743804 | 0 | 24.8953775322 | $S$ |
| 3 | B | 4 | 2.9799883678 | 3.9159049740 | 0.9352934196 | -0.7070086935 | 0.2981079514 | 1.0045476967 | 24.8975943174 | $S$ |
| 4 | B | 5 | 2.9571756627 | 3.9308880115 | 0.9730898238 | -0.7135375880 | 0.2888042290 | 1.0017890027 | 24.9381286473 | $S$ |
| 5 | C | 4 | 3.7283776591 | 2.3647182241 | 2.3647182236 | -0.8436335399 | 0.6657159417 | 0.6657159418 | 24.8967459026 | $S$ |
| 6 | C | 5 | 3.6627045131 | 2.4102817833 | 2.4102817829 | -0.8612089381 | 0.6549877885 | 0.6549877885 | 24.8589057477 | $S$ |

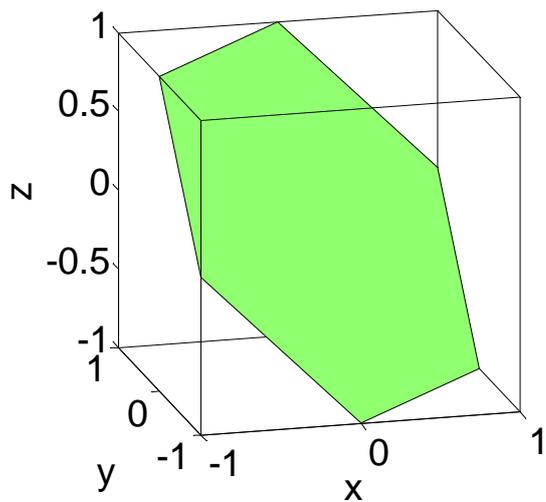

**Fig. 7** One regular Hex cross section of the cube.

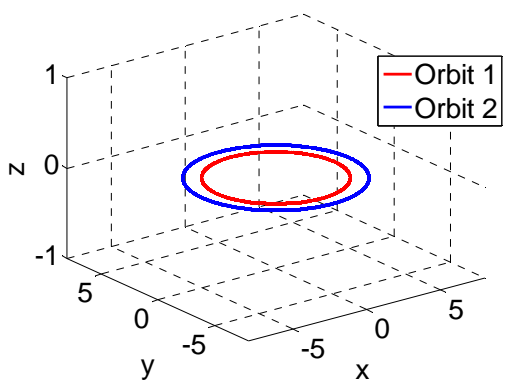

(**a**)

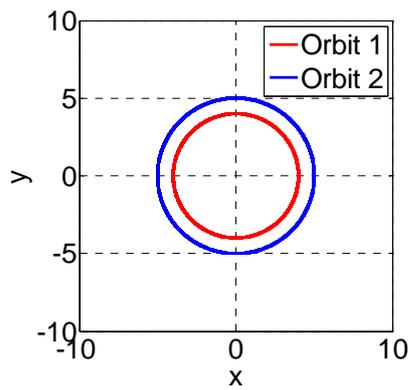

(**b**)

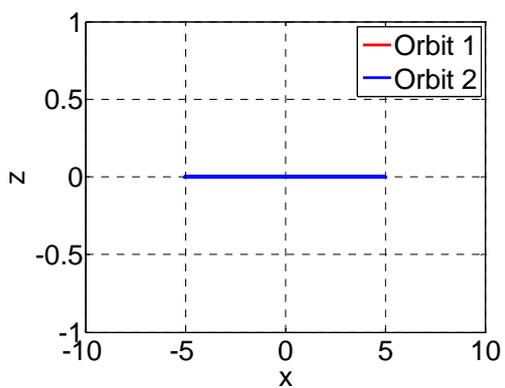

(**c**)

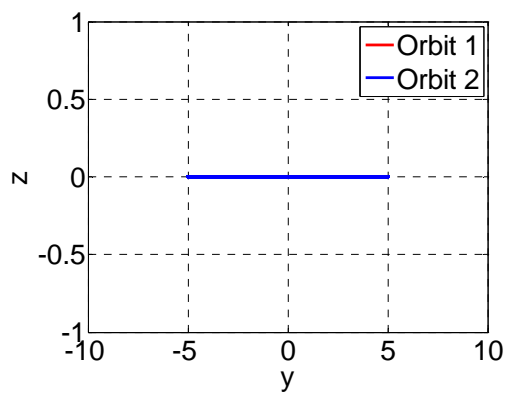

(**d**)

**Fig. 8** Periodic orbits of family A. (**a**) Orbits in the three-dimensional space; (**b**) Projection of orbits onto the *xy*-plane; (**c**) Projection of orbits onto the *xz*-plane; (**d**) Projection of orbits onto the *yz*-plane.

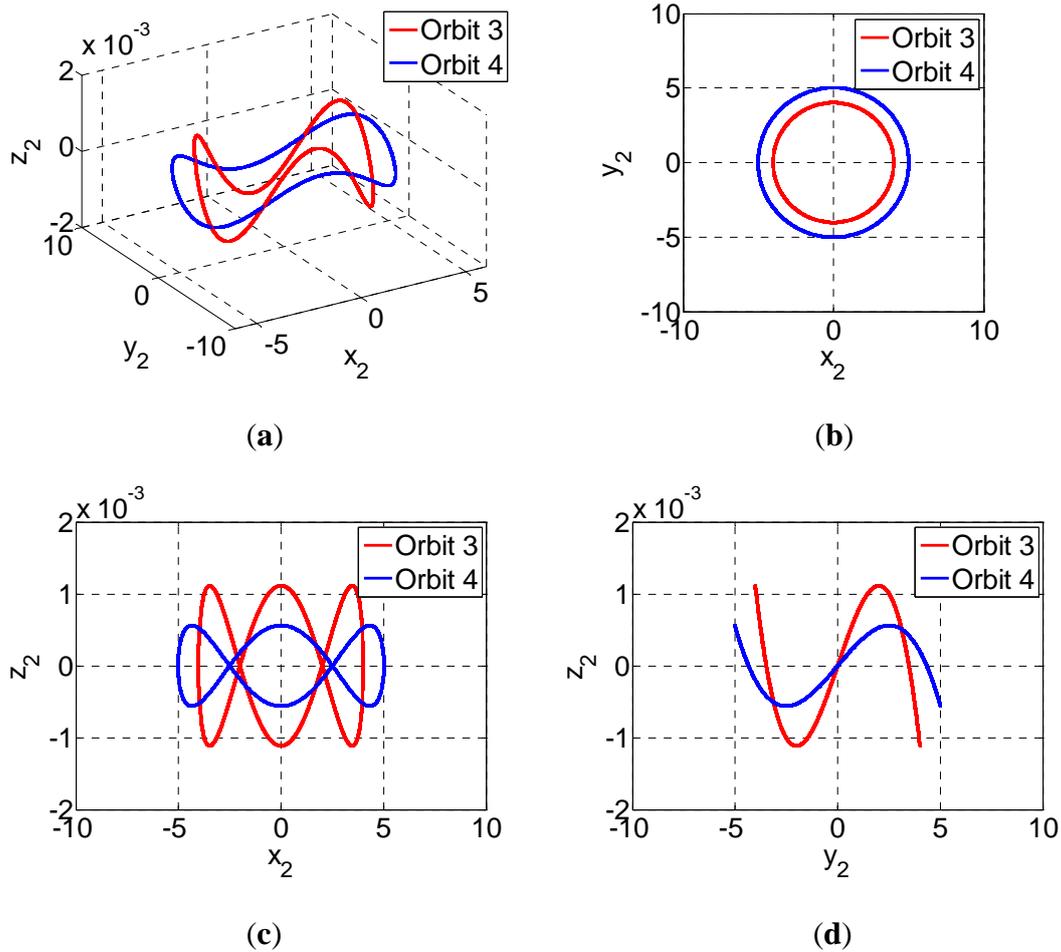

**Fig. 9** Periodic orbits of family B. The Cartesian coordinate system $Ox_2y_2z_2$ is established with the origin $O$ located at the center of the fixed cube and $x_2y_2$-plane coinciding with the plane that contains the regular Hex cross section. (**a**) Orbits in the three-dimensional space; (**b**) Projection of orbits onto the $x_2y_2$-plane; (**c**) Projection of orbits onto the $x_2z_2$-plane; (**d**) Projection of orbits onto the $y_2z_2$-plane.

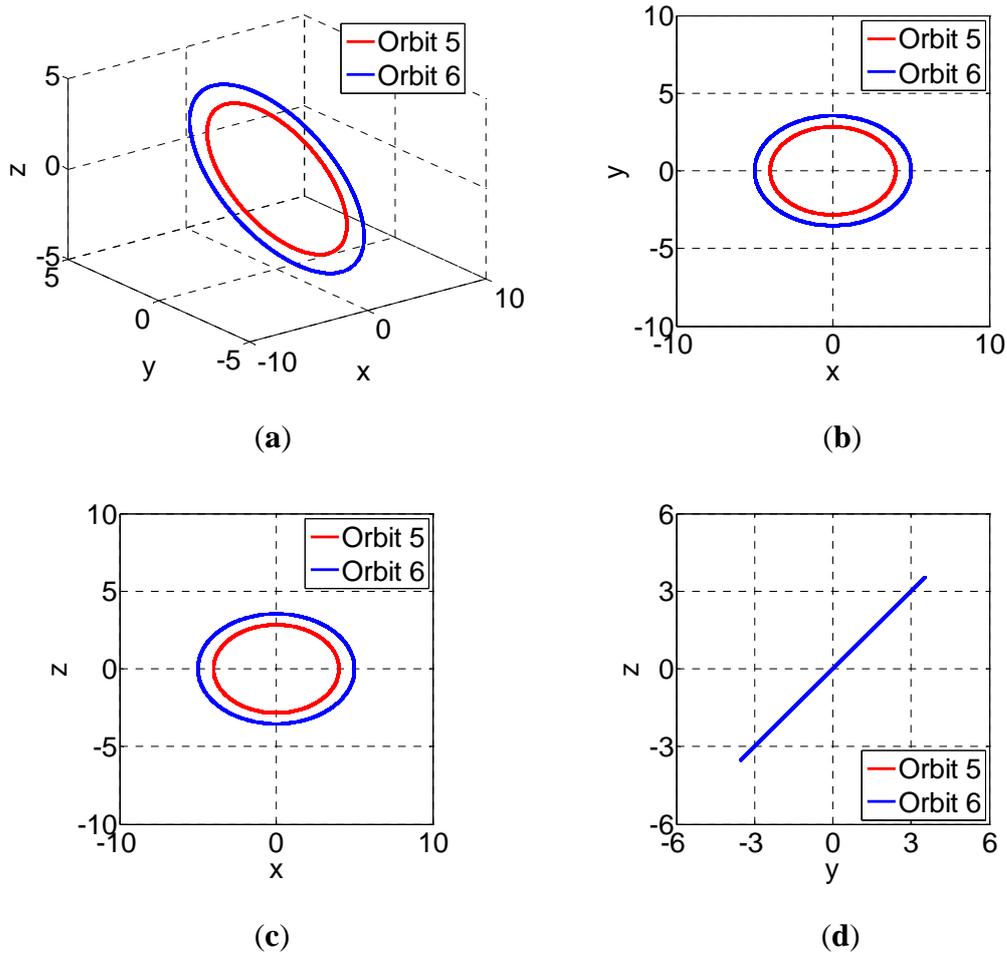

**Fig. 10** Periodic orbits of family C. (**a**) Orbits in the three-dimensional space; (**b**) Projection of orbits onto the *xy*-plane; (**c**) Projection of orbits onto the *xz*-plane; (**d**) Projection of orbits onto the *yz*-plane.

Family A (Fig. 8).

These orbits are located in the *xy*-plane (or other symmetry planes parallel to the face of the cube), and are symmetric with respect to both the *x*-axis and the *y*-axis. It can be seen that these orbits are all quasi-circular. Three pairs of eigenvalues of the monodromy matrix are all located on the unit circle, so these orbits are stable.

Family B (Fig. 9).

These orbits are located near the asymmetry plane that contains the regular Hex cross section of the cube, and are not planar orbits. The regular Hex cross section divides the cube into two equal parts; however, the two parts are asymmetry. It can be seen that the amplitude of motion in $z_2$ direction is three orders of magnitude lower than that in $x_2$ and $y_2$ directions. Three pairs of eigenvalues of the monodromy matrix are all located on the unit circle, so these orbits are stable.

Family C (Fig. 10).

These orbits lie in one diagonal plane that is perpendicular to the $yz$-plane (or other diagonal planes), and are symmetric with respect to the $x$-axis. Three pairs of eigenvalues of the monodromy matrix are all located on the unit circle, so these orbits are stable.

The cube owns three symmetry planes parallel to the face of itself, six symmetry planes in the diagonal plane, and four asymmetry planes that contain the regular Hex cross section. It can be seen that these periodic orbits are located either in the symmetry planes, or near the asymmetry planes that contain the regular Hex cross section. Periodic orbits of these three families are all stable.

## 6. Conclusions

This paper investigates the dynamics of orbits in the gravity field of a fixed homogeneous cube, and finds several types of periodic orbits. Based on the method of the Poincaré surface of section, periodic orbits in the two different types of symmetry planes of the fixed cube are easily found. While in general positions, periodic orbits can be found by the homotopy method. Simulations show that periodic orbits around the cube can be divided into three different families: orbits in the planes parallel to the face of the cube, orbits in the diagonal plane, and orbits in the asymmetry planes that contain the regular Hex cross section. The stability of these periodic orbits is determined based on the eigenvalues of the monodromy matrix. The results demonstrate that periodic orbits of these three families are all stable. The work about dividing irregular shaped celestial bodies into a set of the cubes is in progress.

## Appendix: The potential of the cube

Based on the polyhedral method (Werner and Scheeres 1997), the potential of the fixed cube at a certain point $P(x, y, z)$ in space is calculated as

$$
\begin{aligned}
U = G\sigma \Big\{ &-2(x+a)(z-a)\ln\frac{r_1+r_4+2a}{r_1+r_4-2a} + 2(x+a)(y+a)\ln\frac{r_1+r_5+2a}{r_1+r_5-2a} \\
&-2(x-a)(y+a)\ln\frac{r_2+r_6+2a}{r_2+r_6-2a} + 2(x-a)(y-a)\ln\frac{r_3+r_7+2a}{r_3+r_7-2a} \\
&-2(y+a)(z-a)\ln\frac{r_1+r_2+2a}{r_1+r_2-2a} + 2(x-a)(z-a)\ln\frac{r_2+r_3+2a}{r_2+r_3-2a} \\
&+2(y-a)(z-a)\ln\frac{r_3+r_4+2a}{r_3+r_4-2a} - 2(x+a)(y-a)\ln\frac{r_4+r_8+2a}{r_4+r_8-2a} \\
&+2(y+a)(z+a)\ln\frac{r_5+r_6+2a}{r_5+r_6-2a} - 2(x-a)(z+a)\ln\frac{r_6+r_7+2a}{r_6+r_7-2a} \\
&-2(y-a)(z+a)\ln\frac{r_7+r_8+2a}{r_7+r_8-2a} + 2(x+a)(z+a)\ln\frac{r_5+r_8+2a}{r_5+r_8-2a} \\
&+2(a-z)^2 \left[ \arctan\frac{\mathbf{r}_1 \cdot \mathbf{r}_2 \times \mathbf{r}_3}{r_1 r_2 r_3 + r_1(\mathbf{r}_2 \cdot \mathbf{r}_3) + r_2(\mathbf{r}_1 \cdot \mathbf{r}_3) + r_3(\mathbf{r}_1 \cdot \mathbf{r}_2)} + \arctan\frac{\mathbf{r}_1 \cdot \mathbf{r}_3 \times \mathbf{r}_4}{r_1 r_3 r_4 + r_1(\mathbf{r}_3 \cdot \mathbf{r}_4) + r_3(\mathbf{r}_1 \cdot \mathbf{r}_4) + r_4(\mathbf{r}_1 \cdot \mathbf{r}_3)} \right] \\
&+2(a+z)^2 \left[ \arctan\frac{\mathbf{r}_8 \cdot \mathbf{r}_7 \times \mathbf{r}_6}{r_6 r_7 r_8 + r_6(\mathbf{r}_7 \cdot \mathbf{r}_8) + r_7(\mathbf{r}_6 \cdot \mathbf{r}_8) + r_8(\mathbf{r}_6 \cdot \mathbf{r}_7)} + \arctan\frac{\mathbf{r}_6 \cdot \mathbf{r}_5 \times \mathbf{r}_8}{r_5 r_6 r_8 + r_5(\mathbf{r}_6 \cdot \mathbf{r}_8) + r_6(\mathbf{r}_5 \cdot \mathbf{r}_8) + r_8(\mathbf{r}_6 \cdot \mathbf{r}_5)} \right] \\
&+2(a-x)^2 \left[ \arctan\frac{\mathbf{r}_2 \cdot \mathbf{r}_6 \times \mathbf{r}_3}{r_2 r_3 r_6 + r_2(\mathbf{r}_3 \cdot \mathbf{r}_6) + r_6(\mathbf{r}_2 \cdot \mathbf{r}_3) + r_3(\mathbf{r}_2 \cdot \mathbf{r}_6)} + \arctan\frac{\mathbf{r}_6 \cdot \mathbf{r}_7 \times \mathbf{r}_3}{r_3 r_6 r_7 + r_6(\mathbf{r}_7 \cdot \mathbf{r}_3) + r_7(\mathbf{r}_6 \cdot \mathbf{r}_3) + r_3(\mathbf{r}_6 \cdot \mathbf{r}_7)} \right] \\
&+2(a+x)^2 \left[ \arctan\frac{\mathbf{r}_1 \cdot \mathbf{r}_4 \times \mathbf{r}_5}{r_1 r_4 r_5 + r_1(\mathbf{r}_4 \cdot \mathbf{r}_5) + r_4(\mathbf{r}_1 \cdot \mathbf{r}_5) + r_5(\mathbf{r}_1 \cdot \mathbf{r}_4)} + \arctan\frac{\mathbf{r}_4 \cdot \mathbf{r}_8 \times \mathbf{r}_5}{r_4 r_5 r_8 + r_4(\mathbf{r}_5 \cdot \mathbf{r}_8) + r_8(\mathbf{r}_4 \cdot \mathbf{r}_5) + r_5(\mathbf{r}_4 \cdot \mathbf{r}_8)} \right] \\
&+2(a-y)^2 \left[ \arctan\frac{\mathbf{r}_4 \cdot \mathbf{r}_3 \times \mathbf{r}_7}{r_3 r_4 r_7 + r_3(\mathbf{r}_4 \cdot \mathbf{r}_7) + r_4(\mathbf{r}_3 \cdot \mathbf{r}_7) + r_7(\mathbf{r}_3 \cdot \mathbf{r}_4)} + \arctan\frac{\mathbf{r}_7 \cdot \mathbf{r}_8 \times \mathbf{r}_4}{r_4 r_7 r_8 + r_4(\mathbf{r}_7 \cdot \mathbf{r}_8) + r_7(\mathbf{r}_4 \cdot \mathbf{r}_8) + r_8(\mathbf{r}_4 \cdot \mathbf{r}_7)} \right] \\
&+2(a+y)^2 \left[ \arctan\frac{\mathbf{r}_1 \cdot \mathbf{r}_5 \times \mathbf{r}_6}{r_1 r_5 r_6 + r_1(\mathbf{r}_5 \cdot \mathbf{r}_6) + r_5(\mathbf{r}_1 \cdot \mathbf{r}_6) + r_6(\mathbf{r}_1 \cdot \mathbf{r}_5)} + \arctan\frac{\mathbf{r}_2 \cdot \mathbf{r}_1 \times \mathbf{r}_6}{r_1 r_2 r_6 + r_1(\mathbf{r}_2 \cdot \mathbf{r}_6) + r_2(\mathbf{r}_1 \cdot \mathbf{r}_6) + r_6(\mathbf{r}_1 \cdot \mathbf{r}_2)} \right] \Big\},
\end{aligned}
$$

where the Cartesian coordinate system $Oxyz$ is established with the origin $O$ located at the center of the fixed cube and the three coordinate axes coinciding with the symmetrical axes of the cube; $G$ is the gravitational constant; $\sigma$ is the constant mass density of the cube; $a$ is the half-length of the cubic edge; $\mathbf{r}_i$ ($i=1,2,\ldots,8$) is the vector from the origin $O$ to one of the eight vertices of the cube, and $r_i$ is the norm of $\mathbf{r}_i$.

**Acknowledgements**

This research was supported by the National Natural Science Foundation of China (No. 10832004 and No. 11072122).